\documentclass[a4paper,USenglish,cleveref,autoref,thm-restate]{lipics-v2021}

\usepackage[ruled, vlined, linesnumbered]{algorithm2e}
\usepackage[fixlanguage]{babelbib}
\usepackage{todonotes}
\usepackage{csquotes}
\usepackage{pifont}
\usepackage{ragged2e}

\newcommand{\xmark}{\ding{55}}

\graphicspath{{./graphics/pdf/}}

\title{Experimental Comparison of PC-Trees and PQ-Trees}

\author{Simon D. Fink}{Faculty of Informatics and Mathematics, University of Passau, Germany}{finksim@fim.uni-passau.de}{https://orcid.org/0000-0002-2754-1195}{DFG-grant Ru-1903/3-1}

\author{Matthias Pfretzschner}{Faculty of Informatics and Mathematics, University of Passau, Germany}{pfretzschner@fim.uni-passau.de}{https://orcid.org/0000-0002-5378-1694}{DFG-grant Ru-1903/3-1}

\author{Ignaz Rutter}{Faculty of Informatics and Mathematics, University of Passau, Germany}{rutter@fim.uni-passau.de}{https://orcid.org/0000-0002-3794-4406}{DFG-grant Ru-1903/3-1}

\authorrunning{S.\,D. Fink and M. Pfretzschner and I. Rutter} 

\Copyright{Simon D. Fink and Matthias Pfretzschner and Ignaz Rutter} 



\supplement{\url{https://github.com/N-Coder/pc-tree/}}

\funding{Work partially supported by DFG-grant Ru-1903/3-1.}

\nolinenumbers
\hideLIPIcs  

\EventEditors{John Q. Open and Joan R. Access}
\EventNoEds{2}
\EventLongTitle{42nd Conference on Very Important Topics (CVIT 2016)}
\EventShortTitle{CVIT 2016}
\EventAcronym{CVIT}
\EventYear{2016}
\EventDate{December 24--27, 2016}
\EventLocation{Little Whinging, United Kingdom}
\EventLogo{}
\SeriesVolume{42}
\ArticleNo{23}

\crefname{step}{step}{steps}

\newcommand{\restrPosSet}{\texttt{SER-POS}\xspace}
\newcommand{\restrImSet}{\texttt{SER-IMP}\xspace}
\newcommand{\planSet}{\texttt{DIR-PLAN}\xspace}

\NewDocumentCommand\subplot{O{}m}{\begin{subfigure}[t]{0.5\linewidth}
    \centering
    \includegraphics[width=\linewidth,keepaspectratio]{graphics/plots/#2}
    \vspace{-1cm}
    \caption{#1}
    \label{fig:#2}
  \end{subfigure}}

\newlength{\punctuationfootlength}
\newcommand{\punctuationfootnote}[2]{#2\settowidth{\punctuationfootlength}{#2}\hspace{-0.5\punctuationfootlength}\nolinebreak\footnote{#1}}
\AtBeginDocument{\DeclareFontShape{T1}{lmr}{m}{scit}{<->ssub*lmr/m/scsl}{}}
\begin{document}

\maketitle

\begin{abstract}
  PQ-trees and PC-trees are data structures that represent sets of linear and circular orders, respectively, subject to constraints that specific subsets of elements have to be consecutive.
  While equivalent to each other, PC-trees are conceptually much simpler than PQ-trees; updating a PC-trees so that a set of elements becomes consecutive requires only a single operation, whereas PQ-trees use an update procedure that is described in terms of nine transformation templates that have to be recursively matched and applied.

  Despite these theoretical advantages, to date no practical PC-tree implementation is available.
  This might be due to the original description by Hsu and McConnell~\cite{Hsu2003} in some places only sketching the details of the implementation.
  In this paper, we describe two alternative implementations of PC-trees.
  For the first one, we follow the approach by Hsu and McConnell, filling in the necessary details and also proposing improvements on the original algorithm.
  For the second one, we use a different technique for efficiently representing the tree using a Union-Find data structure.
  In an extensive experimental evaluation we compare our implementations to a variety of other implementations of PQ-trees that are available on the web as part of academic and other software libraries.
  Our results show that both PC-tree implementations beat their closest fully correct competitor, the PQ-tree implementation from the OGDF library~\cite{Chimani2014,Leipert1997}, by a factor of 2 to 4, showing that PC-trees are not only conceptually simpler but also fast in practice.
  Moreover, we find the Union-Find-based implementation, while having a slightly worse asymptotic runtime, to be twice as fast as the one based on the description by Hsu and McConnell.
\end{abstract}

\newpage

\section{Introduction}

PQ-trees represent linear orders of a ground set subject to constraints that require specific subsets of elements to be consecutive.
Similarly, PC-trees do the same for circular orders subject to consecutivity constraints.
PQ-trees were developed by Booth and Lueker~\cite{Booth1976} to solve the consecutive ones problem, which asks whether the columns of a Boolean matrix can be permuted such that the 1s in each row are consecutive.
PC-trees are a more recent generalization introduced by Shih and Hsu~\cite{Shih1999} to solve the circular consecutive ones problem, where the 1s in each row only have to be circularly consecutive.

Though PQ-trees represent linear orders and PC-trees represent circular orders, Haeupler and Tarjan~\cite{Haeupler2008} show that in fact PC-trees and PQ-trees are equivalent, i.e., one can use one of them to implement the other without affecting the asymptotic running time.
The main difference between PQ-trees and PC-trees lies in the update procedure.
The update procedure takes as input a PQ-tree (a PC-tree) $T$ and a subset $U$ of its leaves and produces a new PQ-tree (PC-tree) $T'$ that represents exactly the linear orders (circular orders) represented by $T$ where the leaves in $U$ appear consecutively.
The update procedure for PC-trees consists only of a single operation that is applied independently of the structure of the tree.
In contrast, the update of the PQ-tree is described in terms of a set of nine template transformations that have to be recursively matched and applied.

PQ-trees have numerous applications, e.g., in planarity testing~\cite{Booth1976, Shih1999}, recognition of interval graphs~\cite{Booth1976} and genome sequencing~\cite{Benzer1959}.
Nevertheless, PC-trees have been adopted more widely, e.g., for constrained planarity testing problems~\cite{Blaesius2016,br-pclp-17} due to their simpler update procedure.
Despite their wide applications and frequent use in theoretical algorithms, few PQ-tree implementations and even fewer PC-tree implementations are available.
\Cref{tab:eval-impls} shows an overview of all PC/PQ-tree implementations that we are aware of, though not all of them are working.

In this paper we describe the first correct and generic implementations of PC-trees.
\Cref{sec:prelim} contains an overview of the update procedure for applying a new restriction to a PC-tree.
In \Cref{sec:impl}, we describe the main challenge when implementing PC-trees and how our two implementations take different approaches at solving it.
In \Cref{ch:evaluation}, we present an extensive experimental evaluation, where we compare the performance of our implementations with the implementations of PC-trees and PQ-trees from \Cref{tab:eval-impls}.
Our experiments show that, concerning running time, PC-trees following Hsu and McConnell's original approach beat their closest competitor, the PQ-tree implementation from the OGDF library~\cite{Chimani2014} by roughly a factor 2.
Our second implementation using Union-Find is another 50\% faster than this first one, thus beating the OGDF implementation by a factor of up to 4.

\section{The PC-tree}\label{sec:prelim}

A \emph{PC-tree} $T$ is a tree without degree-2 vertices whose inner nodes are partitioned into \emph{P-nodes} and \emph{C-nodes}.
Edges incident to C-nodes have a circular order that is fixed up to reversal, whereas edges incident to P-nodes can be reordered arbitrarily.
Traversing the tree according to fixed orders around the inner nodes determines a circular ordering of the leaves $L$ of the tree.
Any circular permutation of $L$ that can be obtained from $T$ after arbitrarily reordering the edges around P-nodes and reversing orders around C-nodes is a \emph{valid permutation} of $L$.
In this way a PC-tree represents a set of circular permutations of $L$.

When applying a \emph{restriction} $R\subseteq L$ to $T$, we seek a new tree that represents exactly the valid permutations of $L$ where the leaves in $R$ appear consecutively.
We call a restriction \emph{impossible} if there is no valid permutation of $L$ where the leaves in $R$ are consecutive.
Thus, restriction $R$ is possible if and only if the edges incident to P-nodes can be rearranged and orders of edges incident to C-nodes can be reversed
in such a way that all leaves in $R$ are consecutive.
Updating a PC-tree to enforce the new restriction can thus be done by identifying and adapting the nodes that decide about the consecutivity of the elements of $R$ and then changing the tree to ensure that this consecutivity can no longer be broken.

\begin{figure}[t]
  \begin{subfigure}[t]{0.65\linewidth}
    \centering
    \includegraphics[scale=0.73]{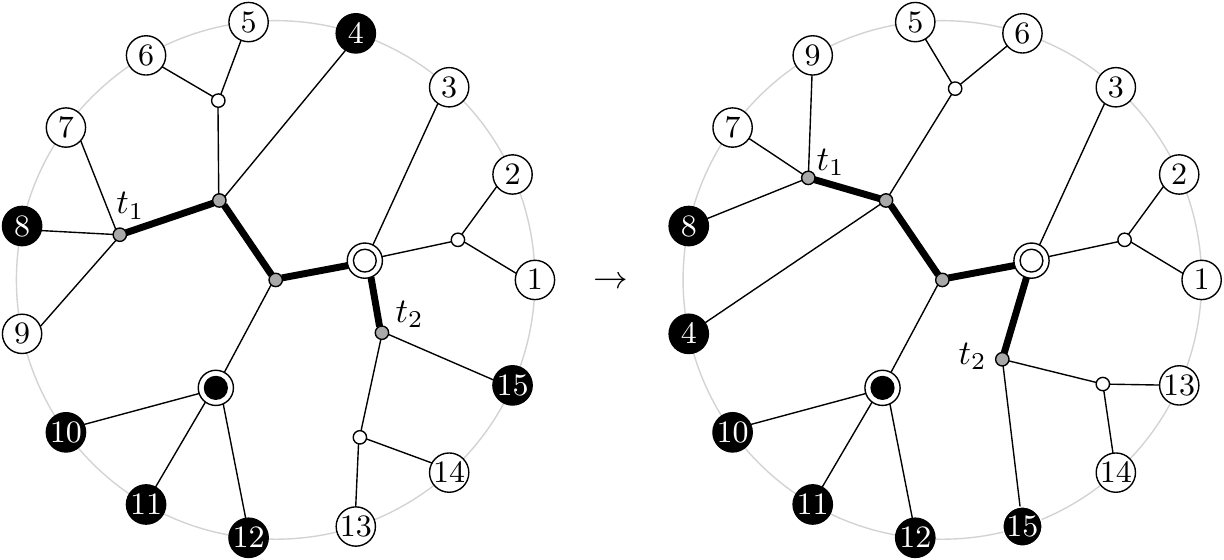}
    \vspace{-1cm}
    \caption{}
    \label{fig:terminalPathPrelim}
  \end{subfigure}\hfill \begin{subfigure}[t]{0.32\linewidth}
    \centering
    \includegraphics[scale=0.73]{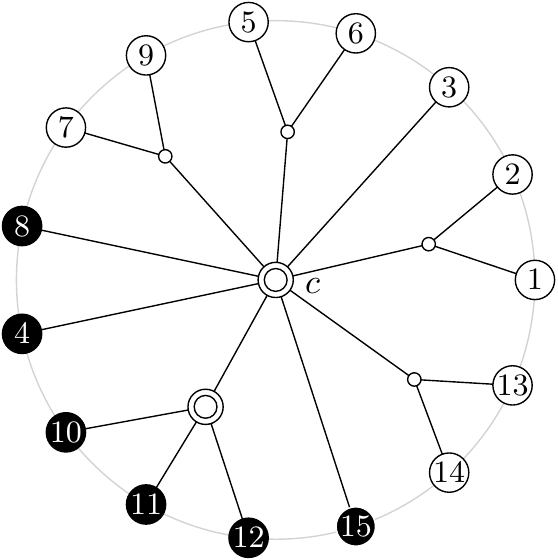}
    \vspace{-0.5cm}
    \caption{}
    \label{fig:updatedTreePrelim}
  \end{subfigure}
  \caption{
    \textbf{(a)} Two equivalent PC-Trees with their nodes colored according to the restriction $\{4,8,10,11,12,15\}$.
    C-nodes are represented by big double circles and the P-nodes are represented by small circles.
    The white nodes represent empty nodes, the black nodes represent full nodes and the gray nodes represent partial nodes.
    The thick edges represent the terminal path with terminal nodes $t_{1}$ and $t_{2}$.
    As the restriction is possible, all full leaves of the tree on the left can be made consecutive, as shown on the right.
    Furthermore all nodes that must be modified lie on a path.\\
    \textbf{(b)} Updated PC-tree with new central C-node $c$.
  }
  \label{fig:updatePrelim}
\end{figure}

Let a leaf $x\in L$ be \emph{full} if $x \in R$ and \emph{empty} otherwise.
We call an edge \emph{terminal} if the two subtrees separated by the edge both contain at least one empty and at least one full leaf.
Exactly the endpoints of all terminal edges need to be ``synchronized'' to ensure that all full leaves are consecutive.
Hsu and McConnell~\cite{Hsu2003,Hsu2004} show that $R$ is possible if and only if the terminal edges form a path and all nodes of this path
can be flipped so that all full leaves are on one side and all empty leaves are on the other.
This path is called the \emph{terminal path}, the two nodes at the ends of the terminal path are the \emph{terminal nodes}. 
Observe that each node in $T$ that is adjacent to two subtrees of which one only contains full leaves and the other contains only empty leaves is contained in the terminal path.
Figure~\ref{fig:terminalPathPrelim} illustrates the terminal path.

\begin{figure}[t]
  \centering
  \includegraphics[scale=0.73]{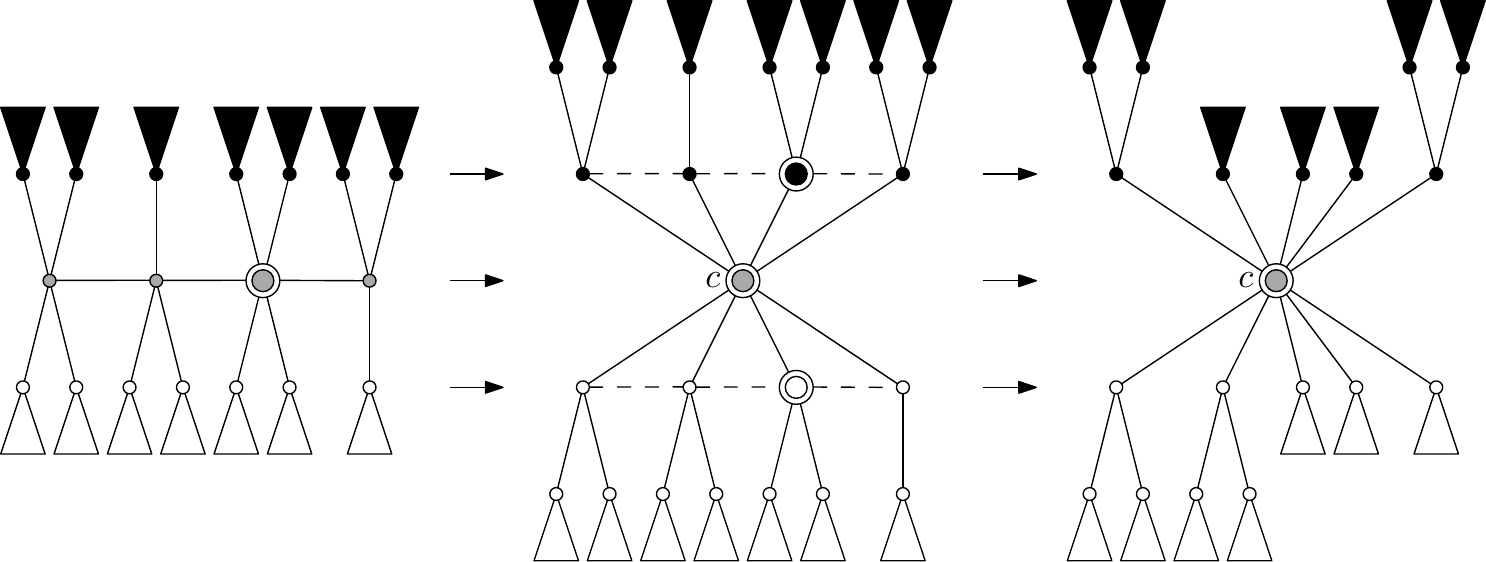}
  \caption{
    Left: The terminal path with all full subtrees shown in black on top and all empty subtrees shown in white on the bottom.
    Middle: The updated PC-tree, where all terminal edges were deleted, all nodes on the terminal were split in a full and empty half and all new nodes were connected to a new C-node $c$.
    Right: The PC-Tree after contracting all new C-nodes and all degree-2 P-nodes into $c$.
  }
  \label{fig:updateStep}
\end{figure}

When updating $T$ in order to apply the restriction, every node on the terminal path is split into two nodes, one of which holds all edges to neighbors of the original node whose subtree has only full leaves, the other holds all edges to empty neighbors, while terminal edges are deleted.
A new central C-node $c$ is created that is adjacent to all the split nodes in such a way that it preserves the order of the neighbors around the terminal path.
Contracting all edges to the split C-nodes incident to $c$ and contracting all nodes with degree two results in the updated tree that represents the new restriction~\cite{Hsu2003,Hsu2004}.
\Cref{fig:updatePrelim} shows an example of this update, while \Cref{fig:updateStep} details changes made to the terminal path.

It remains to efficiently find the terminal edges, and thus the subtrees with mixed full and empty leaves.
To do so, Hsu and McConnell first choose an arbitrary node of the tree as root.
They also assign labels to the inner nodes of the tree, marking an inner node (and conceptually the subtree below it) \emph{partial} if at least one of its neighbors (i.e. children or parent) is full, \emph{full} if all its neighbors except one (which usually is the parent) are full, and \emph{empty} otherwise.
Then, an edge is terminal if and only if it lies on a path between two partial nodes~\cite{Hsu2003,Hsu2004}.
Assigning the labels and subsequently finding the terminal edges can be done by two bottom-up traversals of the tree.
We summarize these steps in the following, more fine-granular description of Hsu and McConnell's algorithm for updating the PC-tree \cite[Algorithm 32.2]{Hsu2004}:
\subparagraph{Algorithm for Applying Restrictions.}\label{main-alg}
To add a new restriction $R$ to a PC-tree $T$:
\begin{enumerate}
\item\label[step]{step:alg-label}    Label all partial and full nodes by searching the tree bottom-up from all full leaves.
\item\label[step]{step:alg-tp}       Find the terminal path by walking the tree upwards from all partial nodes in parallel.
\item\label[step]{step:alg-flip}     Perform flips of C-nodes and modify the cyclic order of edges incident to P-nodes so that all full leaves lie on one side of the path.
\item\label[step]{step:alg-split}    Split each node on the path into two nodes, one incident to all edges to full leaves and one incident to all edges to empty leaves.
\item\label[step]{step:alg-delete}   Delete the edges of the path and replace them with a new C-node $c$, adjacent to all split nodes, whose cyclic order preserves the order of the nodes on this path.
\item\label[step]{step:alg-contract} Contract all edges from $c$ to adjacent C-nodes, and contract any node that has only two neighbors.
\end{enumerate}

\section{Our Implementations}\label{sec:impl}
The main challenge posed to the data structure for representing the PC-tree is that, in \cref{step:alg-contract}, it needs to be able to merge arbitrarily large C-nodes in constant time for the overall algorithm to run in linear time.
This means that, whenever C-nodes are merged, updating the pointer to a persistent C-node object on every incident edge would be too expensive.
Hsu and McConnell (see \cite[Definition 32.1]{Hsu2004}) solve this problem by using C-nodes that, instead of having a permanent node object, are only represented by the doubly-linked list of their incident half-edges, which we call \emph{arcs}.
This complicates various details of the implementation, like finding the parent pointer of a C-node, which are only superficially covered in the initial work of Hsu and McConnell~\cite{Hsu2003}.
These issues are in part remedied by the so called \emph{block-spanning pointers} introduced in the later published book chapter~\cite{Hsu2004}, which are related to the pointer borrowing strategy introduced by Booth and Lueker~\cite{Booth1976}.
These block-spanning pointers link the first and last arc of a consecutive block of full arcs (i.e. the arcs to full neighbors) around a C-node and can be accompanied by temporary C-node objects, see the blue dashed arcs in \Cref{fig:cNodeTPCase-fullI,fig:cNodeTPCase-fullA,fig:fullBlocks} in the Appendix for an example.
Whenever a neighbor of a C-node becomes full, either a new block is created for the corresponding arc of the C-node (\Cref{fig:fullBlocksAppend} left), an adjacent block grows by one arc (\Cref{fig:fullBlocksAppend} right), or the two blocks that now became adjacent are merged (\Cref{fig:fullBlocksMerge}).

Using this data structure, Hsu and McConnell show that the addition of a single new restriction $R$ takes $O(p + |R|)$ time, where $p$ is the length of the terminal path, and that applying restrictions $R_1,\ldots,R_k$ takes $\Theta(|L|+\sum_{i=1}^k|R_i|)$ time~\cite{Hsu2003,Hsu2004}.
Especially for \cref{step:alg-label,step:alg-tp}, they only sketch the details of the implementation, making it hard to directly put it into practice.
In \Cref{sec:algo}, we fill in the necessary details for these steps and also refine their runtime analysis, showing that \cref{step:alg-label} can be done in $O(|R|)$ time and \cref{step:alg-tp} can be done in $O(p)$ time.
Using the original procedures by Hsu and McConnell, \cref{step:alg-flip,step:alg-split} can be done in $O(|R|)$ time and \cref{step:alg-delete,step:alg-contract} can be done in $O(p)$ time.

For our first implementation, which we call HsuPC, we directly implemented these steps in C++, using the data structure without permanent C-node objects as described by Hsu and McConnell.
During the evaluation, we realized that traversals of the tree are expensive.
This is plausible, as they involve a lot of pointer-dereferencing to memory segments that are not necessarily close-by, leading to cache misses.
To avoid additional traversals for clean-up purposes, we store information that is valid only during the update procedure with a timestamp.
Furthermore, we found that keeping separate objects for arcs and nodes and the steps needed to work around the missing C-node objects pose a non-negligible overhead.

To remove this overhead, we created a second version of our implementation, which we call UFPC, using a Union-Find tree for representing C-node objects:
Every C-node is represented by an entry in the Union-Find tree and every incident child edge stores a reference to this entry.
Whenever two C-nodes are merged, we apply \texttt{union} to both entries and only keep the object of the entry that survives.
This leads to every lookup of a parent C-node object taking amortized $O(\alpha(|L|))$ time, where $\alpha$ is the inverse Ackermann function.
Although this makes the overall runtime super-linear, the experimental evaluation following in the next section shows that this actually improves the performance in practice.
As a second change, the UFPC no longer requires separate arc and node objects, allowing us to use a doubly-linked tree consisting entirely of nodes that store pointers to their parent node, left and right sibling node, and first and last child node.
Edges are represented implicitly by the child node whose parent is the other end of the edge.
Note that of the five stored pointers, a lookup in the Union-Find data structure is only needed for resolving the parent of a node.

In \Cref{sec:algo}, we describe our algorithmic improvements and differences of both implementations.
We use the Union-Find data structure from the OGDF~\cite{Chimani2014} and plan to merge our UFPC implementation into the OGDF.
Furthermore, both implementations should also be usable stand-alone with a custom Union-Find implementation.
The source code for both implementations, our evaluation harness and all test data are available on GitHub (see \Cref{tab:eval-impls}).

\section{Evaluation} \label{ch:evaluation}

In this section, we experimentally evaluate our PC-tree implementations by comparing the running time for applying a restriction with that of various PQ- and PC-tree implementations that are publicly available.  In the following we describe our method for generating test cases, our experimental setup and report our results.

\subsection{Test Data Generation}

To generate PQ-trees and restrictions on them, we make use of the planarity test by Booth and Lueker~\cite{Booth1976}, one of the initial applications of PQ-trees.
This test incrementally processes vertices one by one according to an $st$-ordering.
Running the planarity test on a graph with $n$ vertices applies $n-1$ restrictions to PQ-trees of various sizes.
Since not all implementations provide the additional modification operations necessary to implement the planarity test, we rather export, for each step of the planarity test, the current PQ-tree and the restriction that is applied to it as one instance of our test set.
We note that the use of $st$-orderings ensures that the instances do not require the ability of the PC-tree to represent circular permutations, making them good test cases for comparing PC-trees and PQ-trees. 

In this way, we create one test set \restrPosSet consisting of only PQ-trees with possible restrictions by exporting the instances from running the planarity test on a randomly generated biconnected planar graph
for each vertex count $n$ from $1000$ to $20,000$ in steps of $1000$
and each edge count $m \in \{2n,3n-6\}$.
To avoid clutter, to remedy the tendency of the planarity test to produce restrictions with very few leaves, and to avoid bias from trivial optimizations such as filtering trivial restrictions with $|R| \in \{1,|L|-1,|L|\}$, which is present in some of the implementations, we filter test instances where $|R|$ lies outside the interval $[5,|L|-2]$.
Altogether, this test set contains $199,831$ instances, whose distribution with regards to tree and restriction size is shown in \Cref{fig:distplot_full_leaves_tree_size_possible}.

To guard against overly permissive implementations, we also create a small test set \restrImSet of impossible restrictions.
It is generated in the same way, by adding randomly chosen edges to the graphs from above until they become non-planar.
In this case the planarity test fails with an impossible restriction at some point; we include these $3,800$ impossible restrictions in the set, see \Cref{fig:distplot_full_leaves_tree_size_impossible}.

As most of the available implementations have no simple means to store and load a PQ-/PC-tree, we serialize each test instance as a set of restrictions that create the tree, together with the additional new restriction.
When running a test case, we then first apply all the restrictions to reobtain the tree, and then measure the time to apply the new restriction from the test case.
The prefix \texttt{SER-} in the name of both sets emphasizes this serialization.

To be able to conduct a more detailed comparison of the most promising implementations, we also generate a third test set with much larger instances.
As deserializing a PC- or PQ-tree is very time-consuming, we directly use the respective implementations in the planarity test by Booth and Lueker~\cite{Booth1976}, thus calling the set \planSet.
We generated 10 random planar graphs with $n$ vertices and $m$ edges for each $n$ ranging from $100,000$ to $1,000,000$ in steps of $100,000$ and each $m\in\{2n, 3n-6\}$, yielding 200 graphs in total.
The planarity test then yields one possible restriction per node. As we only want to test big restrictions, we filter out restrictions with less than 25 full leaves, resulting in \planSet containing $564,300$ instances.

\begin{figure}[t]
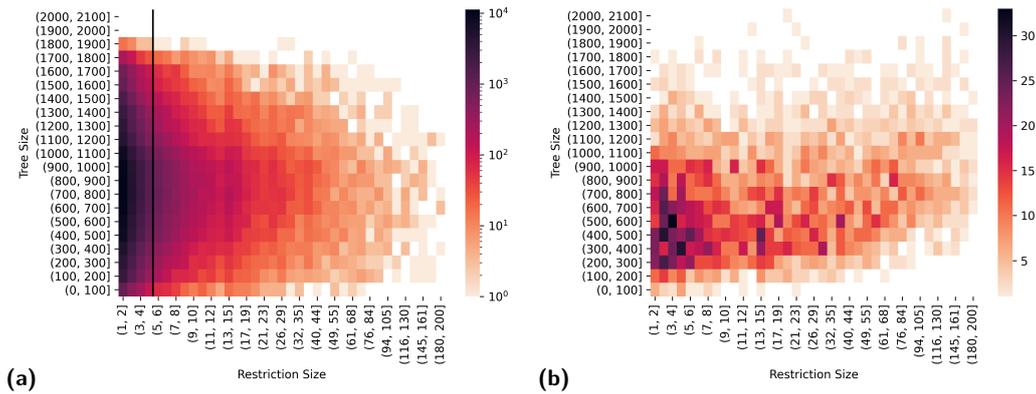

  \centering
  \subplot{distplot_full_leaves_tree_size_possible}\hfill
  \subplot{distplot_full_leaves_tree_size_impossible}
  \caption{
    Distribution of tree and restriction size for the data sets \textbf{(a)} \restrPosSet and \textbf{(b)} \restrImSet.
    Please note the different color scales.
    The \restrPosSet instances that are left of the black line are too small and filtered out.
  }
  \label{fig:distplots}
\end{figure}

\subsection{Experimental Setup}

\newcommand{\PQRTreeF}{PQR-Tree\footnotemark[1]}
\begin{sidewaystable}[p]
  \centering
  \begin{tabular}{l|c|l|c|c|r|p{7cm}}
\textbf{Name} & \textbf{Type} & \textbf{Context}               & \textbf{Language} & \textbf{Correct} & \textbf{Errors} & \textbf{URL}                                                                     \\ \hline
HsuPC                                       & PC-Tree       & our impl., based on \cite{Hsu2004} & C++           &  \checkmark      &   0 & \url{https://github.com/N-Coder/pc-tree/tree/HsuPCSubmodule}                      \\ \hline
UFPC                                        & PC-Tree       & our impl. using Union-Find     & C++               &  \checkmark      &   0 & \url{https://github.com/N-Coder/pc-tree}                                          \\ \hline
Luk\&Zhou                                   & PC-Tree       & student course project         & C++               &  $-$             & $-$ & \url{https://github.com/kwmichaelluk/pc-tree}                                     \\ \hline
Hsu           \cite{Hsu2003a}               & PC-Tree       & planarity test prototype       & C++               &  n.a.            & $-$ & \url{http://qa.iis.sinica.edu.tw/graphtheory}                                     \\ \hline
Noma          \cite{Boyer2004}              & PC-Tree       & planarity test evaluation      & C++               &  n.a.            & $-$ & \url{https://www.ime.usp.br/~noma/sh}                                             \\ \hline
OGDF          \cite{Leipert1997}            & PQ-Tree       & planarity testing              & C++               &  \checkmark      &   0 & \url{https://ogdf.github.io}                                                      \\ \hline
Gregable                                    & PQ-Tree       & biclustering                   & C++               &  \checkmark      &   0 & \url{https://gregable.com/2008/11/pq-tree-algorithm.html}                         \\ \hline
BiVoC         \cite{Grothaus2006}           & PQ-Tree       & automatic layout of biclusters & C++               &  \xmark          &  71 & \url{https://bioinformatics.cs.vt.edu/~murali/papers/BiVoC}                       \\ \hline
Reisle                                      & PQ-Tree       & student project                & C++               &  \xmark          & 236 & \url{https://github.com/creisle/pq-trees}                                         \\ \hline
GraphSet      \cite{EstrellaBalderrama2009} & PQ-Tree       & visual graph editor            & C++               &  \xmark          & 580 & \url{http://graphset.cs.arizona.edu}                                              \\ \hline
Zanetti       \cite{Zanetti2012}            & \PQRTreeF     & extension of PQ-Trees          & Java              &  \xmark          & 454 & \url{https://github.com/jppzanetti/PQRTree}                                       \\ \hline
CppZanetti                                  & \PQRTreeF     & our C++ conversion of Zanetti  & C++               &  \xmark          & 454 & \url{https://github.com/N-Coder/pc-tree#installation}                             \\ \hline
JGraphEd      \cite{Harris2004}             & PQ-Tree       & visual graph editor            & Java              &  \xmark          &  11 & \url{https://www3.cs.stonybrook.edu/~algorith/implement/jgraphed/implement.shtml} \\ \hline
GTea          \cite{Cregten2017}            & PQ-Tree       & visual graph theory tool       & Java              &  $-$             & $-$ & \url{https://github.com/rostam/GTea}                                              \\ \hline
TryAlgo                                     & PQ-Tree       & consecutive-ones testing       & Python            &  $-$             & $-$ & \url{https://tryalgo.org/en/data structures/2017/12/15/pq-trees}                  \\ \hline
SageMath                                    & PQ-Tree       & interval graph detection       & Python            &  \checkmark      &   0 & \url{https://doc.sagemath.org/html/en/reference/graphs/sage/graphs/pq_trees.html} \\
  \end{tabular}
  \caption{
    Implementations considered for the evaluation.   Implementations that are entirely unusable as they are incomplete or crash/produce incorrect results on almost all inputs (marked with $-$) and those where no stand-alone PC-/PQ-tree implementation could be extracted (marked with n.a.) could not be evaluated.  Correct implementations are marked with~\checkmark and implementations that are functional, but do not always produce correct results are marked with \xmark.  These two categories are included in our experimental evaluation. The last column shows how many of $203,630$ restrictions in the sets \restrPosSet and \restrImSet failed.
  }
  \label{tab:eval-impls}

  \justify
  \noindent\rule{5cm}{0.4pt}\medskip

  \noindent
  \begin{minipage}[t]{.013\textwidth}
    \footnotemark[1]
  \end{minipage}\begin{minipage}[t]{.987\textwidth}
    \footnotesize PQR-Trees are a variant of PQ-Trees that can also represent impossible restrictions, replacing any node that would make a restriction impossible by an R-node (again allowing arbitrary permutation).
    To make the implementations comparable, we abort early whenever an impossible restriction is detected and an R-node would be generated.
  \end{minipage}
\end{sidewaystable}
\stepcounter{footnote}

\Cref{tab:eval-impls} gives an overview of all implementations we are aware of,
although not all implementations could be considered for the evaluation.

The three existing implementations of PC-trees we found are incomplete and unusable (Luk\&Zhou) or tightly intertwined with a planarity test in such a way that we were not able to extract a generic implementation of PC-trees (Hsu, Noma).
We further exclude two PQ-tree implementations as they either crash or produce incorrect results on almost all inputs (GTea) or have an excessively poor running time (TryAlgo).
Among the remaining PQ-tree implementations only three correctly handle all our test cases (OGDF, Gregable, SageMath).
Several other implementations have smaller correctness issues:
After applying a fix to prevent segmentation faults in a large number of cases for BiVoC, the remaining implementations crash (BiVoC, GraphSet, Zanetti) and/or produce incorrect results (Reisle, JGraphEd, Zanetti) on a small fraction of our tests; compare the last column of \Cref{tab:eval-impls}.
We nevertheless include them in our evaluation.

We changed the data structure responsible for mapping the input to the leaves of the tree for BiVoC and Gregable from \texttt{std::map} to \texttt{std::vector} to make them competitive.
Moreover, BiVoC, Gregable and GraphSet use a rather expensive cleanup step that has to be executed after each update operation.
As this could probably largely be avoided by the use of timestamps, we do not include the cleanup time in their reported running times.
For SageMath the initial implementation turned out to be quadratic, which we improved to linear by removing unnecessary recursion.
As Zanetti turned out to be a close competitor to our implementation in terms of running time, we converted the original Java implementation to C++ to allow a fair comparison.
This decreased the runtime by one third while still producing the exact same results.
All other non-C++ implementations were much slower or had other issues, making a direct comparison of their running times within the same language environment as our implementations unnecessary.
Further details on the implementations can be found in~\Cref{sec:impl-details}.

Each experiment was run on a single core of a Intel Xeon E5-2690v2 CPU (3.00 GHz, 10 Cores) with 64 GiB of RAM, running Linux Kernel version 4.19.
Implementations in C++ were compiled with GCC 8.3.0 and optimization \texttt{-O3 -march=native -mtune=native}.
Java implementations were executed on OpenJDK 64-Bit Server VM 11.0.9.1 and Python implementations were run with CPython 3.7.3.
For the Java implementations we ran each experiment several times, only measuring the last one to remove startup-effects and to facilitate optimization by the JIT compiler.
We used OGDF version 2020.02 (Catalpa) to generate the test graphs.

\subsection{Results}

\begin{figure}[p]
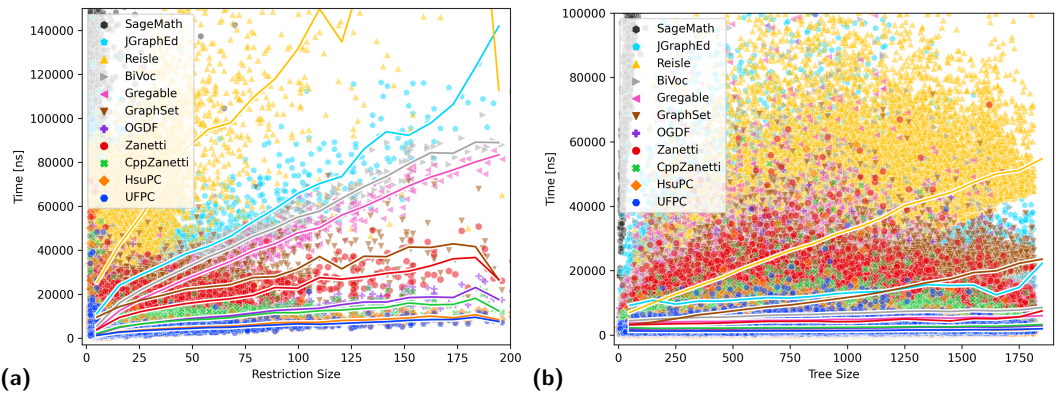

  \centering
  \subplot{relplot_time_full_leaves_possible}\hfill
  \subplot{relplot_time_tree_size_possible}
  \caption{Runtime for \restrPosSet restrictions depending on \textbf{(a)} restriction size and \textbf{(b)} tree size.}
  \label{fig:relplot_possible}
\end{figure}

\begin{figure}[p]
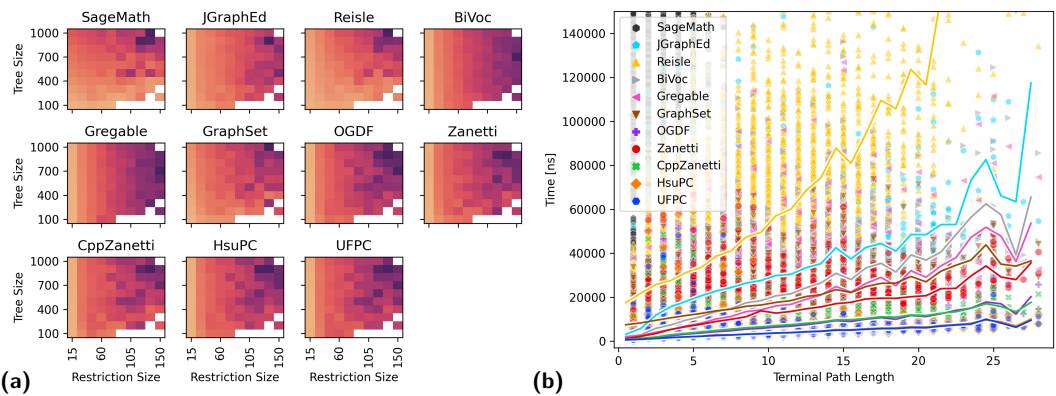

  \centering
  \subplot{heatmap_time_possible_buckets}\hfill
  \subplot{relplot_time_tp_length_possible}
  \caption{
    \textbf{(a)} A heatmap showing the average runtime of \restrPosSet restrictions, depending on both the size of the restriction and the size of the tree.
    The color scale is based on the maximum runtime of each respective implementation.
   \textbf{(b)} Runtime for \restrPosSet restrictions depending on the terminal path length.
  }
\end{figure}

\begin{figure}[p]
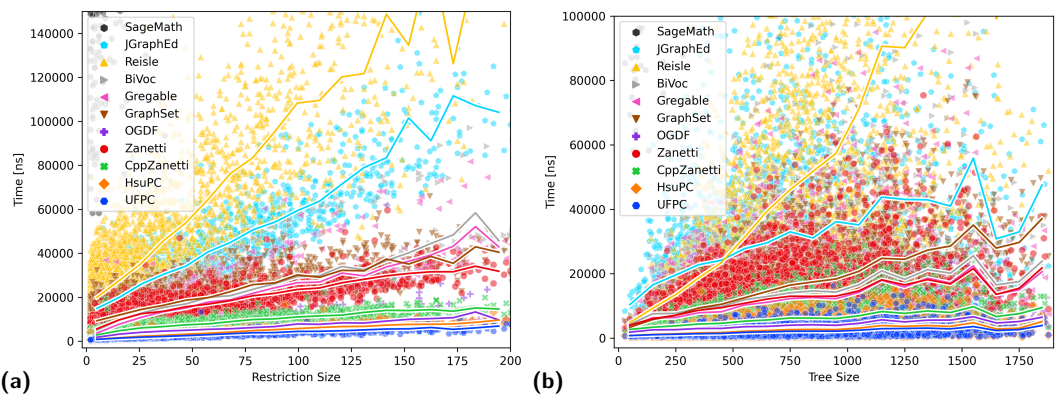

  \centering
  \subplot{relplot_time_full_leaves_impossible}\hfill
  \subplot{relplot_time_tree_size_impossible}
  \caption{
   Runtime for \restrImSet restrictions depending on \textbf{(a)} restriction size and \textbf{(b)} tree size for all implementations.
  }
  \label{fig:relplot_impossible}
\end{figure}

Our experiments turn out that SageMath, even with the improvements mentioned above, is on average 30 to 100 times slower than all other implementations.\footnote{Part of this might be due to the overhead of running the code with CPython. As the following analysis shows, SageMath also has other issues, allowing us to safely exclude it.}
For the sake of readability, we scale our plots to focus on the other implementations.
As the main application of PC-/PQ-trees is applying possible restrictions, we first evaluate on the dataset \restrPosSet.
\Cref{fig:relplot_possible} shows the runtime for individual restrictions based on the size of the restriction (i.e. the number of full leaves) and the overall size of the tree.
\Cref{fig:relplot_time_full_leaves_possible} clearly shows that for all implementations the runtime is linear in the size of the restriction.
\Cref{fig:relplot_time_tree_size_possible} suggests that the runtime of Reisle and GraphSet does not solely depend on the restriction size, but also on the size of the tree.
To verify this, we created for each implementation a heatmap that indicates the average runtime depending on both the tree size and the restriction size, shown in \Cref{fig:heatmap_time_possible_buckets}.
The diagonal pattern shown by SageMath, Reisle, and GraphSet confirms the dependency on the tree size.
All other implementations exhibit vertical stripes, which shows that their runtime does not depend on the tree size.
Finally, \Cref{fig:relplot_time_tp_length_possible} shows the runtime compared to the terminal path length.
As expected, all implementations show a linear dependency on the terminal path length, with comparable results to \Cref{fig:relplot_time_full_leaves_possible}.

\begin{figure}[t]
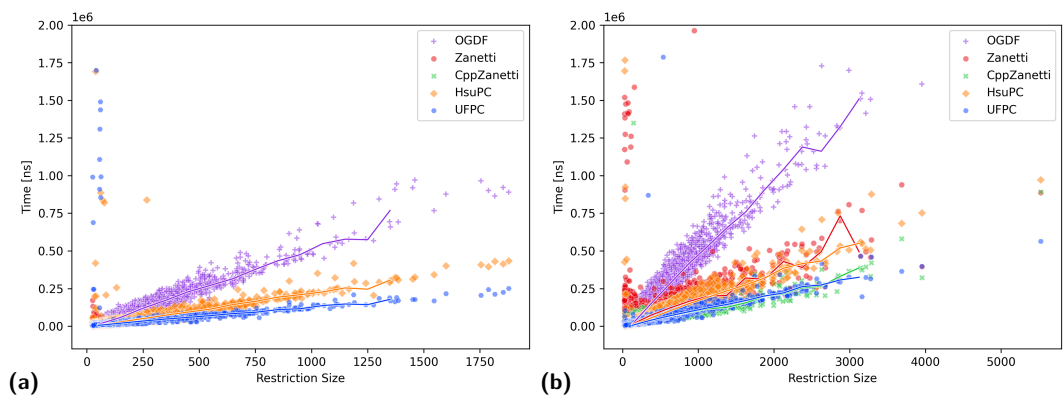

  \centering
  \subplot{planaritytest_scatter_size_2n}\hfill
  \subplot{planaritytest_scatter_size_3n}
  \caption{
  	Runtime of individual restrictions of \planSet with OGDF, Zanetti and our implementations for graphs of size \textbf{(a)} $m=2n$ and \textbf{(b)} $m=3n-6$.
  }
  \label{fig:planaritytest_scatter_size}
\end{figure}

\Cref{fig:relplot_impossible} shows the performance on the dataset \restrImSet.
The performance is comparable with that on \restrPosSet.
Noteworthy is that Zanetti performs quite a bit worse, which is due to its implementation not being able to detect failure during a labeling step.  It always performs updates until a  so-called R-node would be generated.
Altogether, the data from \restrPosSet and \restrImSet shows that the implementations GraphSet, OGDF, Zanetti, HsuPC and UFPC are clearly superior to the others.
In the following, we conduct a more detailed comparison of these implementations by integrating them into a planarity test and running them on much larger instances, i.e., the data set \planSet.
In addition to an update method, this requires a method for replacing the now-consecutive leaves by a P-node with a given number of child leaves.
Adding the necessary functionality would be a major effort for most of the implementations, which is why we only adapted the most efficient implementations to run this set.
We also exclude GraphSet from this experiment; the fact that it scales linearly with the tree size causes the planarity test to run in quadratic time (see also \Cref{sec:impl-details}).
\Cref{fig:planaritytest_scatter_size} again shows the runtime of individual restrictions depending on the restriction size.
Curiously, Zanetti produces incorrect results for nearly all graphs with $m=2n$ in \Cref{fig:planaritytest_scatter_size_2n}.
As the initial tests already showed, the implementation has multiple flaws; one major issue is already described in an issue on GitHub, while we give a small example of another independent error in \Cref{fig:zanettiBroken} in the appendix.
Both plots show that HsuPC is more than twice as fast as OGDF and that UFPC is again close to two times faster than HsuPC.
Zanetti's runtime is roughly the same as that of HsuPC, while converting its Java code to C++ brings the runtime down close to that of UFPC.

As OGDF is the slowest, we use it as baseline to calculate the speedup of the other implementations.
\Cref{fig:planaritytest_speedup_size} shows that the runtime improvement for all three implementations is the smallest for small restrictions,
quickly increasing to the final values of roughly $0.4$ times the runtime of OGDF for HsuPC and $0.25$ for both CppZanetti and UFPC.
\Cref{fig:planaritytest_speedup_tp_length} shows the speedup depending on the length of the terminal path.
For very short terminal paths (which are common in our datasets, see \Cref{fig:planaritytest_distplot_size_tp_length} in the appendix), both implementations are again close; but already for slightly longer terminal paths UFPC quickly speeds up to being roughly 20\% faster than CppZanetti.
This might be because creating the central node in \cref{step:alg-delete} is more complicated for UFPC,
as the data structure without edge objects does not allow arbitrarily adding and removing edges (which is easier for HsuPC)
and allowing circular restrictions forces UFPC to also pay attention to various special cases (which are not necessary for PQ-trees).

\begin{figure}[t]
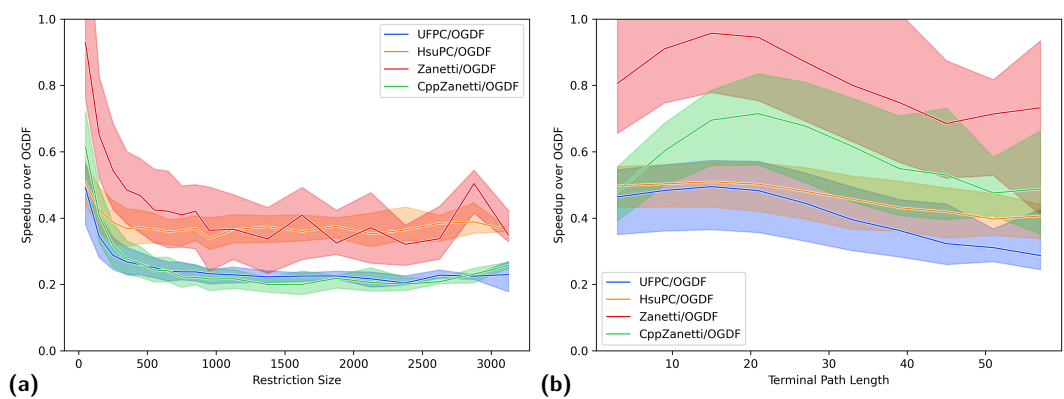

  \centering
  \subplot{planaritytest_speedup_size}\hfill
  \subplot{planaritytest_speedup_tp_length}
  \caption{
    Median performance increase depending on \textbf{(a)} the size of the restriction and \textbf{(b)} the terminal path length, with OGDF as baseline. The shaded areas show the interquartile range.
  }
  \label{fig:planaritytest_speedup}
\end{figure}

\section{Conclusion} \label{ch:conclusion}
In this paper we have presented the first fully generic and correct implementations of PC-trees.
One implementation follows the original description of Hsu and McConnell~\cite{Hsu2003,Hsu2004}, which contains several subtle mistakes in the description of the labeling and the computation of the terminal path.
This may be the reason why no fully generic implementation has been available so far.
A corrected version that also includes several small simplifications is given in the appendix.

Furthermore, we provided a second, alternative implementation, using Union-Find to replace many of the complications of Hsu and McConnell's original approach.
Technically, this increases the runtime to $O((|R|+p)\cdot\alpha(|L|))$, where $\alpha$ is the inverse Ackerman function.
In contrast, our evaluations show that the Union-Find-based approach is even faster in practice, despite the worse asymptotic runtime.

Our experimental evaluation with a variety of other implementations reveals that surprisingly few of them seem to be fully correct.
Only three other implementation have correctly handled all our test cases.
The fastest of them is the PQ-tree implementation of OGDF, which our Union-Find-based PC-tree implementation beats by roughly a factor of 4.
Interestingly, the Java implementation of PQR-trees by Zanetti achieves a similar speedup once ported to C++.
However, Zanetti's Java implementation is far from correct and it is hard to say whether it is possible to fix it without compromising its performance.

Altogether, our results show that PC-trees are not only conceptually simpler than PQ-trees but also perform well in practice, especially when combined with Union-Find.
To put the speedup of factor 4 into context, we compared the OGDF implementations of the planarity test by Booth and Lueker and the one by Boyer and Myrvold on our graph instances.
The Boyer and Myrvold implementation was roughly 40\% faster than the one by Booth and Lueker.
Replacing the PQ-trees, which are the core part of Booth and Lueker's algorithm, by an implementation that is 4 time faster, might make this planarity test run faster than the one by Boyer and Myrvold.
We leave a detailed evaluation, also taking into account the embedding generation, which our PC-tree based planarity test not yet provides, for future work.

\bibliography{references}

\appendix

\begin{figure}[h]
  \centering
  \includegraphics[width=.5\linewidth,keepaspectratio]{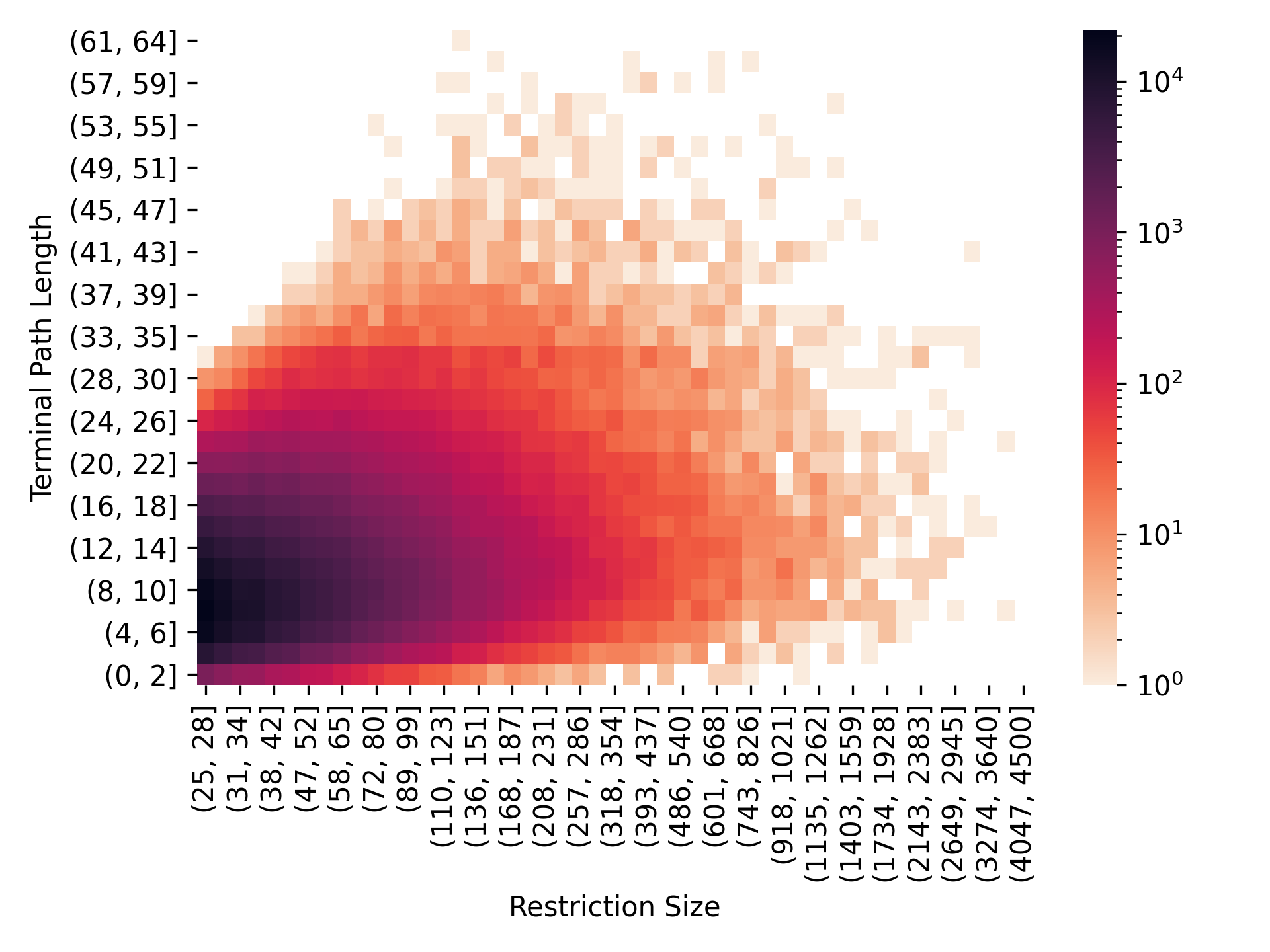}
  \caption{Distribution of tree and restriction size for \planSet.}
  \label{fig:planaritytest_distplot_size_tp_length}
\end{figure}

\section{Details About Evaluated Implementations}\label{sec:impl-details}
\begin{figure}[t]
  \centering
  \begin{subfigure}[t]{0.49\linewidth}
  	\centering
  	\includegraphics[page=1]{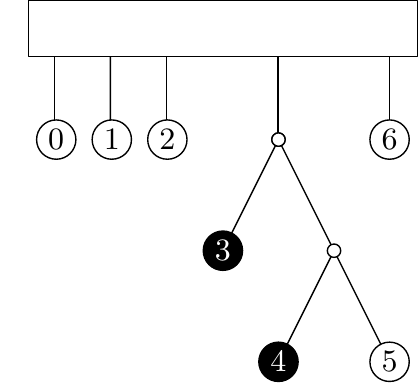}
  \end{subfigure}\hfill
  \begin{subfigure}[t]{0.49\linewidth}
  	\centering
  	\includegraphics[page=2]{graphics/figures/zanetti_broken}
  \end{subfigure}
  \caption{
  	Left: The PQR-tree \texttt{[0 1 2 (3 (4 5)) 6]} not containing any R-nodes with its root Q-node depicted as rectangle and the two P-nodes depicted as small circles.
  	Right: The result of Zanetti's implementation applying the restriction \texttt{\{3,4\}} to the former tree, the tree \texttt{[[3 4 5] 0 1 2 6]} which clearly represents a different set of restrictions.
  }
  \label{fig:zanettiBroken}
\end{figure}

\begin{description}

\item[BiVoC, Gregable]
In the implementations of \emph{BiVoC} and \emph{Gregable}, we improved the mapping from the input to the tree's leaves by replacing \texttt{std::map} with \texttt{std::vector}, as suggested in the code's comments.
As a result, this mapping now takes constant time.
The \texttt{Bubble} method of BiVoC caused segmentation faults due to undefined behavior, because a set iterator is dereferenced and incremented after its corresponding element has been removed. We resolved this issue for our evaluation.
Still, the method \texttt{qNextChild} of BiVoC rarely caused program hangs due to undefined behavior, when the past-the-end iterator of an empty set is incremented.
In the Gregable repository, the author notes that the code ``is known to be buggy on some rare inputs.
A believed to be correct, but harder to use version of this code can be found as a library within BiVoC''.
However, in contrast to BiVoC, Gregable's implementation always produced the expected results in our evaluation.

\item[GraphSet]
In the implementation of \emph{GraphSet}, we removed the entanglement with Microsoft Foundation Classes by replacing its data structures with their corresponding variants from the standard library.
We were unable to get GraphSet's \texttt{Bubble} method to work for our tests.
Instead, we used the approach from their quadratic-time variant of Booth and Lueker's planarity test, where they traverse the entire tree before each reduction in order to find and prepare the pertinent subtree.
Still, GraphSet produced segmentation faults due to null pointer dereferencing in Template Q3.

\item[TryAlgo]
In June 2020, the authors of the \emph{TryAlgo} implementation noted on their website that they ``have problems implementing this data structure, and cannot provide at this point a correct implementation in tryalgo''.
Furthermore, they note that ``the current implementation has a complexity in the order of $n*m$, however an implementation in $O(n+m+s)$ is possible''.
As we thus assumed their implementation to be neither correct nor linear-time, we excluded it from our evaluation.

\item[SageMath]
The main routine \texttt{set\_contiguous} of the PQ-tree of \emph{SageMath} recursively traverses the tree starting from its root as follows:
it first calls \texttt{set\_contiguous} recursively on all children of the current node, then calls \texttt{flatten}, calls \texttt{set\_contiguous} recursively on all children again
and then proceeds to sort the children depending on whether they are full, partial, or empty.
The \texttt{flatten} function for removing degree-2 nodes is implemented to recurse itself on all children in the subtree,
making the runtime of \texttt{set\_contiguous} quadratic in the tree size.
We modified the implementation to only flatten the current level and dropped the second recursive call to \texttt{set\_contiguous},
improving the runtime to linear in the tree size without generating incorrect results.

\item[Zanetti]
We found that \emph{Zanetti}'s data structures became inconsistent after some restrictions, which was also already independently reported on GitHub\punctuationfootnote{\url{https://github.com/jppzanetti/PQRTree/issues/2}}.
This happened mostly after restrictions having a terminal path length greater than 1.
As the restrictions generated when serializing a PC-tree only have very short terminal paths and the inconsistency is usually only found when modifying the same area of the tree again, only few of these cases surfaced in our tests on \restrPosSet.
Only when applying multiple bigger restrictions consecutively, these issues surfaced more often, i.e. at some point during the planarity test for close to all graphs with $m=2n$ and also some of the graphs with $m=3n-6$.
We also found a second, independent issue, where Zanetti's implementation generates C-nodes with their children in the wrong order.
An examples where this happens is shown in \Cref{fig:zanettiBroken}.

As Zanetti's Java implementation still has a very good runtime in practice, we decided to port its Java code to C++ to be able to perform a direct comparison with the other C++ implementations.
As the implementation uses close to none Java specific features, the conversion mostly involved replacing Java Object variables with C pointers and Java utility classes with their C++ stdlib equivalents.
The only non-trivial change was that, because Zanetti stores the Union-Find information directly in the nodes and not in an external array, we had to implement reference counting for Zanetti's tree nodes to ensure that the lifetime of nodes which are no longer part of the tree, but still referenced in the Union-Find data structure, is handled properly.
We made sure that both the Java and the C++ version not only produced equivalent output, but actually keep the same PQR-tree state in memory.
Where both implementations differ is that Java immediately reports inconsistencies of the data structure, e.g. by throwing a \texttt{NullPointerException}, while the \texttt{SIGSEGFAULT} of C++ might not be immediately triggered.
This generates a few more data points with an invalid result, where the Java implementation already crashed.

\end{description}

\section{Algorithmic Improvements}\label{sec:algo}

In this section, we describe further details in which our implementations differ from the description given by Hsu and McConnell and explain the corrections needed for a working implementation.
\Cref{sec:step-label} describes the labeling procedure used by HsuPC and UFPC.
Note that the technical complications involving the arcs and block-spanning pointers only concern the missing C-node objects of HsuPC.
UFPC uses direct references to the concerned nodes instead of arcs and doesn't maintain block-spanning pointers, which is why those parts are not relevant for our second implementation, which is otherwise based on HsuPC.
The same holds for \Cref{sec:step-tp}, where we give a corrected algorithm for enumerating the terminal path.
\Cref{sec:step-tp-impossible} then describes the generic steps needed to detect impossible restrictions.
Lastly, \Cref{sec:step-delete-contract-uf} explains the differing update step of UFPC, while the update step of HsuPC follows Hsu and McConnells original description.

\subsection{Efficiently Labeling and Finding Partial Nodes}\label{sec:step-label}
In our description of the labeling step, we follow the general procedure of Hsu and McConnell~\cite{Hsu2004}, using a bottom-up traversal of the tree, starting at the full leaves.
To implement this, we keep a queue of unprocessed non-full arcs of full nodes, which is initialized with the parent arcs of all full leaves.
If the front of the queue knows its node object, this has to be a P-node and we can simply append the incoming arc to the node's list of full children.
Recall that each full node has exactly one non-full neighbor.
Thus, if this list reaches a size one smaller than the node's degree, we enqueue the parent pointer of the node, if it exists and the parent is not yet full.
The other case, when the now-full node is the root or when the parent has become full before its child, is missing in the description by Hsu and McConnell.
Here, we need to search all children for the single non-full node, and queue this arc instead.
This case is rare and the search of all full children can still be done in time linear in the number of full leaves, thus not affecting the overall runtime.

\begin{figure}[t]
  \centering
  \begin{subfigure}[t]{0.49\linewidth}
    \includegraphics[page=2,scale=0.7]{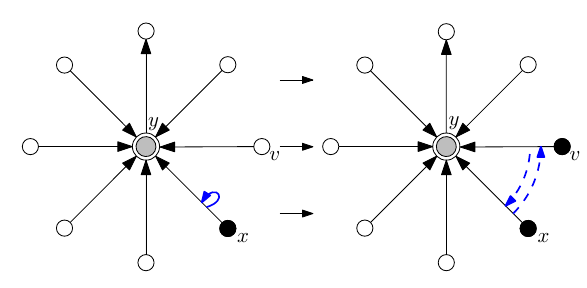}
    \vspace{-1cm}
    \caption{}
    \label{fig:fullBlocksAppend}
  \end{subfigure}\hfill
  \begin{subfigure}[t]{0.49\linewidth}
    \includegraphics[page=4,scale=0.7]{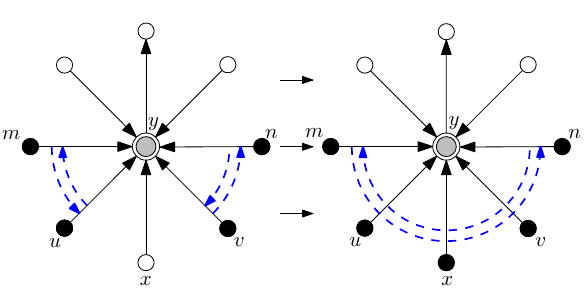}
    \vspace{-1cm}
    \caption{}
    \label{fig:fullBlocksMerge}
  \end{subfigure}
  \caption{\textbf{(a)} A newly created block $x$ of size one growing by one as the neighboring arc $v$ becomes full and is appended. 
    \textbf{(b)} Two blocks that share a common neighbor $x$ being merged once $x$ becomes full.
    The block-spanning pointers are shown as blue, dashed half-arcs.
  }
  \label{fig:fullBlocks}
\end{figure}

If the arc does not point to a permanent (P-)node, we need to maintain (i.e. create/append/merge) the block-spanning pointers around the respective C-node $x$.
The merging of full blocks is illustrated in \Cref{fig:fullBlocks}, see the book chapter by Hsu and McConnell~\cite{Hsu2004} for more details.
If the new endpoints of the C-node's full block now share a common non-full neighbor, the C-node is full and we queue this neighbor $z$.
Note that similar to the case of P-nodes, this node is most often but not necessarily (i.e. for the root) the parent arc.
Still, there is no explicit search required, as we already know the arc to the only non-full neighbor.

In their definition of the data structure Hsu and McConnell note that ``[n]o two C nodes are adjacent, so each of these edges [incident to a C-node] has one end that identifies a neighbor of the C node, and another end that indicates that the end is incident to a C node, without identifying the C node''~\cite[page 32-10]{Hsu2004}.
Considering the PC-tree with 6 leaves representing the restrictions $\{\{1,2\}, \{2,3\}, \{4,5\}, \{5,6\}\}$, which consists of two adjacent C-nodes and (up to reversal and circular shifting) only has the two valid permutations $[1,2,3,4,5,6]$ and $[1,2,3,6,5,4]$, this statement is obviously untrue.
Hsu and McConnell use this property within their planarity test and when they test whether the endpoints of a full block are adjacent to the same arc:
``[i]f $x$ passes this test, it is full, and the full-neighbor counter of $z$ is incremented''~\cite[page 32-11]{Hsu2004}.
According to their argumentation, $z$ has to be a P-node as no two C nodes are adjacent, which is incorrect.
Still, the important information is not the type of the nodes, but that the neighbor $x$ of non-full node $z$ became full,
and our queue-based approach also handles the case of a chain of multiple C-nodes becoming full consecutively.

\subsection{Efficiently Finding the Terminal Path}\label{sec:step-tp}
\begin{figure}[t]
  \centering
  \begin{subfigure}[t]{0.45\linewidth}
    \centering
    \includegraphics[page=5]{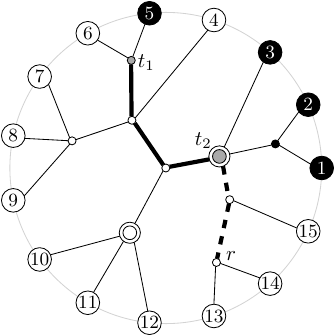}
    \vspace{-0.5cm}
    \caption{}
    \label{fig:tpIshaped}
  \end{subfigure}\hfill \begin{subfigure}[t]{0.45\linewidth}
    \centering
    \includegraphics[page=3]{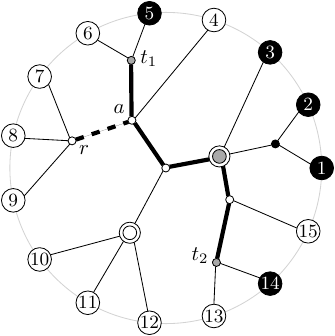}
    \vspace{-0.5cm}
    \caption{}
    \label{fig:tpAshaped}
  \end{subfigure}
  \caption{
    Two examples of a PC-tree with root $r$.
    \textbf{(a)} An I-shaped terminal path where $t_{2}$ is both apex and terminal node.
    \textbf{(b)} An A-shaped terminal path where two ascending paths join in the apex $a$.
  }
\end{figure}

There are two possible structures for the terminal path, assuming the restriction is possible.
Let the \emph{apex} be the highest node on the terminal path, i.e., the node that is an ancestor of all other nodes on the terminal path.
The two cases can now be differentiated based on the position of the apex, which in turn depends on the position of the root node:

\begin{description}
\item[I-Shaped:] If the apex lies on one of the ends of the terminal path and is therefore a terminal node at the same time, the terminal path extends from the other endpoint of the path upwards to the apex, as shown in Figure~\ref{fig:tpIshaped}.
In this case, every node on the terminal path has exactly one child on the terminal path, except for the lower terminal node $t_{1}$.
This also covers the special case where the terminal path has length 0 and there is only a single terminal node $t_{1}=t_{2}$.

\item[A-Shaped:] If the apex does not lie on one of the ends of the terminal path, two ascending paths join in the apex, as shown in Figure~\ref{fig:tpAshaped}.
In this case, the apex $a$ has two children on the terminal path, the terminal nodes $t_{1}$ and $t_{2}$ have none,
and all other nodes on the terminal path have exactly one child that is also on the terminal path.
\end{description}

Hsu and McConnell show that an edge is terminal if and only if it lies on a path in the tree between two partial nodes.
This allows them to conduct parallel searches, starting at every partial node and extending ascending paths through their ancestors at the same rate~\cite{Hsu2004}.
Whenever an already processed node is encountered, expansion of the current path is stopped and the path is instead merged into the path of the already processed node.
This bottom-up traversal can easily be done for P-nodes, but again as C-nodes have no object registered with their incident arcs (i.e. except for a temporary one stored at the two endpoints of their full block), finding their parent arc can be difficult.

\begin{figure}[t]
  \centering
  \begin{subfigure}[t]{0.24\linewidth}
    \centering
    \includegraphics[page=1,width=\linewidth]{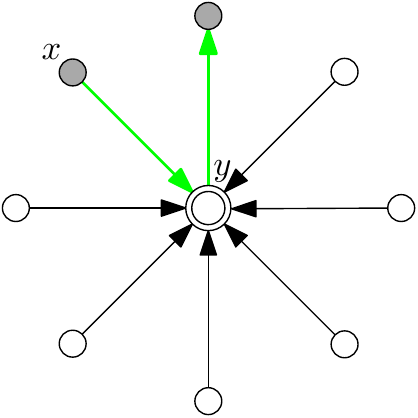}
    \vspace{-0.5cm}
    \caption{An \emph{empty} C-node that is part of a possibly \hbox{\emph{I-shaped}} terminal path.}
    \label{fig:cNodeTPCase-emptyI}
  \end{subfigure}\hfill \begin{subfigure}[t]{0.24\linewidth}
    \centering
    \includegraphics[page=2,width=\linewidth]{graphics/c_node_tp_cases}
    \vspace{-0.5cm}
    \caption{An \emph{empty} C-node that is \emph{apex} of an  \hbox{\emph{A-shaped}} terminal path.}
    \label{fig:cNodeTPCase-emptyA}
  \end{subfigure}\hfill \begin{subfigure}[t]{0.24\linewidth}
    \centering
    \includegraphics[page=3,width=\linewidth]{graphics/c_node_tp_cases}
    \vspace{-0.5cm}
    \caption{A  \emph{full}  C-node that is part of a possibly \hbox{\emph{I-shaped}} terminal path.}
    \label{fig:cNodeTPCase-fullI}
  \end{subfigure}\hfill \begin{subfigure}[t]{0.24\linewidth}
    \centering
    \includegraphics[page=4,width=\linewidth]{graphics/c_node_tp_cases}
    \vspace{-0.5cm}
    \caption{A  \emph{full}  C-node that is \emph{apex} of an  \hbox{\emph{A-shaped}} terminal path.}
    \label{fig:cNodeTPCase-fullA}
  \end{subfigure}\hfill \caption{
Different cases of the terminal path crossing a C-node.
    Empty, partial, and full nodes are drawn in white, gray, and black, respectively.
    The edges are oriented towards the root node.
    The green edges are part of the terminal path, the blue dashed half-arcs depict the block-spanning pointers.
    In cases (c) and (d), the node can be the final node of the terminal path, i.e. a terminal node.
    Thus there might zero, one or two incident terminal edges. 
  }
\end{figure}

Hsu and McConnell distinguish two cases:
Either they arrived at the C-node $x$ via an arc from a partial or empty node, or via an arc from a full node.
In the former case, they ``look at the two neighbors of the child edge in the cyclic order, and one of them must be the parent edge. This takes $O(1)$ time.''
In the latter case, they ``cycle clockwise and counterclockwise through the edges to full neighbors. Of the first edges clockwise and counter-clockwise that are not
to full neighbors, one of these is the parent. In this case, the cost of finding the parent is proportional to the number of full neighbors $x$''~\cite[page 32-12]{Hsu2004}.
Their procedure for the first case is incorrect, as $x$ could also be the apex where both incident terminal arcs are adjacent to each other,
but not to the parent arc (note that if the apex $x$ is also the root, it doesn't even have a parent arc; see \Cref{fig:cNodeTPCase-emptyA}).
The procedure for the second case is also incorrect, as $x$ could again be the apex, where the parent arc (if it exists) is not necessarily adjacent to the full block, as shown in \Cref{fig:cNodeTPCase-fullA}.
Moreover, we can never arrive at $x$ via an arc from a full node, as full nodes cannot be part of the terminal path.

To correctly implement this step, we again use a queue of unprocessed arcs that may be part of the terminal path.
We initialize the queue with the parent arcs of all partial nodes found in the previous step, which is easy to do for partial P-nodes.
To find the parent arc of a C-node $x$, only knowing the two endpoints of a full block from the labeling step, we fix the second case of Hsu and McConnell's case distinction as follows:
If one of the two non-full arcs adjacent to the full block is indeed the parent arc $p$ (see \Cref{fig:cNodeTPCase-fullI}), we do simply continue the path in that direction by queuing $p$.
Otherwise, we store $x$ as apex candidate as we do not need to ascend any further.
This holds for the case when the parent arc $p$ is within the full block, as a full node can not be part of the terminal path.
If $p$ is located somewhere else in the empty block, continuing the terminal path via that node would make it impossible to later split the C-node $x$ into an (exclusively) empty and an (exclusively) full half.

We now process the queue step by step, seeking the parent arc $p$ of the node $x$ to which the removed arc $a$ points.
When we encounter the root, we store it as apex candidate and continue with the next entry of the queue.
We also never process a node twice, proceeding to the next entry if we detect this case.
The hard part now is again finding the parent pointer of a C-node $x$.
We first check whether the arcs next to the incoming arc $a$ are part of a full block.
If both are part of the same block, we can stop ascending as we ran into a full node (the same holds for full P-nodes).
If both are part of two different full blocks, we abort and report an impossible restriction, as there is no flip of $x$ that makes all full leaves consecutive.
If only one is part of a full block, we can use the procedure for initialization described in the previous paragraph to find the parent arc.

Now consider the case where there is no full block adjacent to $a$, i.e. the former case of Hsu and McConnells case distinction.
If the parent arc $p$ is actually adjacent to the incoming arc $a$, we follow that path and queue $p$ (see \Cref{fig:cNodeTPCase-emptyI}).
Otherwise, the terminal path may continue as shown in \Cref{fig:cNodeTPCase-emptyA}, where one of the arcs adjacent to $a$ also belongs to the terminal path and the current node $x$ is the apex.
As we don't know which of the neighboring edges is terminal, we mark $a$ and store $x$ (identified by its incident arc $a$) as the apex candidate.
We can resolve this situation, identify the second predecessor of $x$ and fix $x$ as actual apex if we later process an arc that is adjacent to the accordingly marked arc $a$.

Observe that for each processed arc, we now either found a following parent arc or marked the corresponding node as apex candidate.
If we marked a node as apex candidate and encounter it a second time when ascending from a different node, we can be sure that the terminal path is A-shaped and the current node actually is the apex.
Otherwise, we can't be sure that the apex candidate is the actual apex if it is empty, as we might be extending an I-shaped path of empty nodes above the actual apex.
Thus, if the apex candidate is empty and has only one predecessor, we find the actual apex by ``backtracking'' to the next partial node below the apex candidate.

Furthermore, if there is only one element left in the queue and we haven't found an apex (candidate) yet, all parts of the terminal path have already converged into a single ascending I-shaped path (instead of having met in an A-shape and stopped at the apex) and we are again extending the terminal path above the previously unidentified apex.
We stop processing the queue and backtrack to the next partial node below the last remaining element of the queue.
Otherwise, the queue processing also terminates if an impossible restriction is detected (see \Cref{sec:step-tp-impossible}) or the queue runs empty, already having identified the actual apex.

The backtracking needed to find the actual apex can be done in constant time by storing, for each empty node on the terminal path, the highest partial node in its subtree.
Special care needs to be taken as an empty apex candidate may collide with a later-found actual apex.
In this case, we again extended an I-shaped path of empty nodes above the previously unidentified apex and only later found the second terminal path child of the apex, indicating that the terminal path is actually A-shaped.
As long as both the apex and the apex candidate have the same node as highest partial node in their subtree, the restriction is valid and the apex candidate can simply be discarded.

Observe that the number of arcs on the terminal path is proportional to the length $p$ of the terminal path.
Furthermore, we only check up to two neighbors of each arc and any apex candidate requiring backtracking is at most $p$ nodes above the actual apex.
Thus, the overall runtime of our search for the terminal path is in $O(p)$.
Note that this slightly refines the analysis of Hsu and McConnell~\cite{Hsu2004}, who sometimes scan the full children of a node and thus have a runtime in $O(p+|R|)$.

\subsection{Efficiently Detecting Impossible Restrictions}\label{sec:step-tp-impossible}
Recall that a restriction is possible if and only if all terminal edges form a path
and all nodes on the terminal path can be flipped or rearranged such that their empty and full children are consecutive while separated by the terminal edges.
The only way to violate the first property is when a node has more than two incident terminal edges.
P-nodes can detect directly when this case occurs, while a C-node with more than two incident terminal edges either leads to multiple apex candidates or detects the usage of one of its arc in multiple paths.
As P-nodes allow arbitrary arrangements of their children, only C-nodes can violate the second property
by either having multiple distinct full blocks,
by having a full block that is not adjacent to all incident terminal edges, or
by having no full block and two non-adjacent terminal edges.
All three cases lead to a disconnected terminal-path and thus multiple apex (candidate) nodes, if the impossible restriction is not detected right away because an arc would be part of two different paths.
Thus, we know that the restriction is impossible if and only if we either find a second, incompatible apex (candidate) or a node has more than two incident terminal edges.

\subsection{Deletion and Contraction}\label{sec:step-delete-contract-uf}
Deleting and inserting new edges is simple when using the arc-based tree representation described by Hsu and McConnell.
When using a doubly-linked tree structure similar like the one used by UFPC, no explicit edge objects exists and they are instead encoded by the child-parent relationship of the nodes.
This means that in \cref{step:alg-delete,step:alg-contract} of the update procedure, the child-parent relationship needs to be set immediately and correctly for every change and cannot easily be updated later, as done by Hsu and McConnell.
Thus, UFPC uses a different approach for these two steps:
First, the central node is created and we make sure that it already has its final neighbors. 
This is trivial if the apex is a C-node and thus can simply be reused as is.
Otherwise, we create a new C-node and add up to 4 neighbors:
the apex' first child on the terminal path, a newly created P-node that was reassigned as parent of all full children of the apex,
the second child on the terminal path, and finally the apex with all its empty children remaining.
Note that not all four neighbors might exist or be required, e.g. when the apex has no full or empty children.
Furthermore, the root of the tree is either among the full or the empty children and thus the node that is still connected to the root needs to be installed as parent of the new central node.
Second, we iteratively contract a child of the central node that is part of terminal path into the central node.
C-nodes can again be simply merged, while a P-node $x$ needs to be split into a full and empty node.
P-node $x$ is then replaced by the full node and the empty node with the other terminal path neighbor of $x$ in between, if the latter exists.

\end{document}